# Cassini UVIS Observations of the Io Plasma Torus: II. Radial Variations

Andrew J. Steffl

Laboratory for Atmospheric and Space Physics, 392 UCB, University of Colorado, Boulder, Colorado 80309

E-mail: steffl@colorado.edu

Fran Bagenal

Laboratory for Atmospheric and Space Physics, 392 UCB, University of Colorado, Boulder, Colorado 80309

A. Ian F. Stewart

Laboratory for Atmospheric and Space Physics, 1234 Innovation Drive, Boulder, Colorado 80309-7814

Pages: 50

Tables: 1

Figures: 10

**Proposed Running Head:** Io Torus Radial Variations

**Editorial correspondence to:**

Mr. Andrew J. Steffl

LASP

392 UCB

Boulder, CO 80309-0392

Phone: 303-492-3617

Fax: 303-492-6946

Email: steffl@colorado.edu

**ABSTRACT**.

On January 14, 2001, shortly after the Cassini spacecraft's closest approach to Jupiter, the Ultraviolet Imaging Spectrometer (UVIS) made a radial scan through the midnight sector of Io plasma torus. The Io torus has not been previously observed at this local time. The UVIS data consist of 2-D spectrally dispersed images of the Io plasma torus in the wavelength range of 561Å-1912Å. We developed a spectral emissions model that incorporates the latest atomic physics data contained in the CHIANTI database in order to derive the composition of the torus plasma as a function of radial distance. Electron temperatures derived from the UVIS torus spectra are generally less than those observed during the Voyager era. We find the torus ion composition derived from the UVIS spectra to be significantly different from the composition during the Voyager era. Notably, the torus contains substantially less oxygen, with a total oxygen-to-sulfur ion ratio of 0.9. The average ion charge state has increased to 1.7. We detect S V in the Io torus at the 3σ level. S V has a mixing ratio of 0.5%. The spectral emission model used in can approximate the effects of a non-thermal distribution of electrons. The ion composition derived using a kappa distribution of electrons is identical to that derived using a Maxwellian electron distribution; however, the kappa distribution model requires a higher electron column density to match the observed brightness of the spectra. The derived value of the kappa parameter decreases with radial distance and is consistent with the value of κ=2.4 at 8 $R_J$ derived by the Ulysses URAP instrument (Meyer-Vernet *et al.*, 1995). The observed radial profile of electron column density is consistent with a flux tube content, $NL^2$, that is proportional to $r^{-2}$.

**Key Words:** Jupiter, Magnetosphere; Ultraviolet Observations; Spectroscopy; Io

I. INTRODUCTION

The Io plasma torus is a dense (~2000 cm$^{-3}$) ring of electrons and sulfur and oxygen ions trapped in Jupiter's strong magnetic field, produced by the ionization of ~1 ton per second of neutral material from Io's atmosphere. In situ measurements of the Io plasma torus from the Voyager and Galileo spacecrafts and remote sensing observations from the ground and from space-based UV telescopes have characterized the density, temperature and composition of the plasma as well as the basic spatial structure (see review by Thomas *et al.*, 2004), Extensive measurements of torus emissions made by the Ultraviolet Imaging Spectrograph on the Cassini spacecraft as it flew past Jupiter on its way to Saturn allow us to further examine the spatial and temporal structure of the plasma torus.

On ionization, fresh ions tap the rotational energy of Jupiter (to which they are coupled by the magnetic field). Much of the torus thermal energy is radiated as intense (~10$^{12}$ watts) EUV emissions. The ~100 eV temperature of the torus ions indicates that they have lost more than half of their initial pick-up energy. Electrons, on the other hand, have very little energy at the time of ionization and gain thermal energy from collisions with the ions (as well as through other plasma processes) while losing energy via the EUV emissions that they excite. As a result, the torus electrons have an average thermal energy of ~5 eV, although in situ measurements indicate that the velocity distribution of the torus electrons has a supra-thermal tail (Sittler and Strobel 1985, Frank and Paterson 2000, Meyer-Vernet *et al.* 1995).

Analysis of torus emissions provides estimates of plasma density, composition and temperature (Brown *et al.* 1983). Models of mass and energy flow through the torus can then be used to derive plasma properties such as source strength, source composition, and radial transport timescale (see review by Thomas *et al.* 2004, Delamere and Bagenal 2003, Lichtenberg 2001, Schreier *et al*. 1998). Thus, one aims to relate observations of spatial and temporal variations in torus emissions to the underlying sources, losses and transport processes. Towards this ultimate goal, we present an analysis of observations of the Io torus made by the Cassini spacecraft's Ultraviolet Imaging Spectrograph (UVIS) on January 14, 2001, with emphasis on determining the radial structure. In a companion paper, (Steffl et al.2004), hereafter referred to as paper I, we present examples of the EUV spectra of the torus and its temporal variability as observed during the full 6-month encounter period. Analysis of the temporal structure of the torus is presented in Delamere *et al*. 2004.

## II. UVIS DATA

UVIS consists of two independent, but coaligned, spectrographs: one optimized for the extreme ultraviolet (EUV), which covers a wavelength range of 561Å to 1181Å and the other optimized for the far ultraviolet (FUV), which covers a wavelength range of 1140Å to 1913Å (McClintock *et al*. 1993, Esposito *et al.* 1998, Esposito *et al.* 2000). Each spectrograph is equipped with a 1024 x 64 pixel imaging microchannel plate detector. UVIS pixels are rectangular, and subtend an angle of 1 milliradian in the spatial dimension (i.e. along the length of the slit) and 0.25 milliradians long in the spectral dimension (i.e. along the dispersion direction). Images are obtained of UV-emitting targets with a spectral resolution of ~3Å FWHM, roughly a factor of ten increase in resolution over previous UV spectrographs sent to Jupiter. The spectral range and resolution of the instrument and the extended observation period resulted in the creation of a unique and rich dataset of the Io plasma torus in the extreme and far ultraviolet.

### A. Observations

The data used in this analysis were obtained from a single observational sequence that began at 12:04:25 UT on January 14, 2001. This data represents only a small fraction of the total observations of the Io plasma torus made by UVIS; a general summary of the UVIS Jupiter encounter dataset can be found in paper I. The data consist of 39 spectrally-dispersed images of the Io torus, each with an integration time of 1,000 seconds. During the observation period, Io moved from near western elongation to 0.4 $R_J$ (Jovian radii) east of the planet, as seen from Cassini. By maintaining a constant angular offset of 21.1 milliradians from Io, the center of the UVIS field of view was scanned radially inwards from 10.4 to 4.3 $R_J$. A listing of observational parameters for the data analyzed in this paper is provided in Table 1.

[TABLE 1]

Cassini's closest approach to Jupiter occurred on December 30, 2000, so the spacecraft was well within the dusk sector when it made these observations. The local time of the observed torus ansa was approximately 01:50, i.e. nearly two hours past midnight. This region of local time cannot be seen from earth and was not well observed by either of the Voyager spacecraft.

In contrast to the vast majority of the UVIS observations of the Io torus, the data used in this paper were obtained with the long axis of the UVIS entrance slits oriented approximately perpendicular to Jupiter's rotational equator. The low-resolution slit was used for both channels. This slit has an angular width, as seen from the detector, of 2 milliradians for the EUV channel and 1.5 milliradians for the FUV channel. At the time, Cassini was 244 $R_J$ from Jupiter, resulting in a field of view 0.48 $R_J$ wide for the EUV channel and 0.36 $R_J$ wide for the FUV channel.

In addition to the difference in slit width, there is a small pointing offset between the EUV and FUV channels such that, in this configuration, the fields of view of the two channels do not overlap. The sense of this offset is such that the FUV channel views a section of the torus at a greater radial distance from Jupiter than that viewed by the EUV channel. Fortuitously, the field of view of the FUV channel at a given time lies completely within the field of view of the EUV channel one integration time earlier. Therefore, we have excluded from our analysis the first image from the FUV channel and used the second FUV image in conjunction with the first EUV channel image. Likewise, the third FUV image is used in conjunction with the second EUV image, and so on for the rest of the dataset. Since the two channels view the same radial distance 1,000 seconds apart, systematic errors will be introduced if there are strong longitudinal or temporal variations in torus properties. However, since the integration period less than 3% of Jupiter's rotation period, these systematic effects should be relatively minor. The projection of the UVIS EUV slit field-of-view and the positions of Io and Europa relative to Jupiter are shown in Fig. 1.

[FIGURE 1]

**B. Data Reduction**

The data were reduced and calibrated using techniques similar to those described in greater detail in paper I. To summarize this procedure: the background is subtracted from the raw data, a flat-field correction is applied, and the data are divided by the effective area curve of the instrument to convert it from counts to physical units. Figure 2 shows a calibrated, background-subtracted sample of the data.

**[FIGURE 2]**

To increase signal-to-noise in the data, each 2-D spectral image was averaged over the latitudinal extent of the Io torus (i.e. in the vertical direction on the detector) to create a 1-D spectrum. This was accomplished by first summing along the rows of the detector (the spectral direction) to create a latitudinal emission profile. A Gaussian plus quadratic background was fit to the latitudinal emission profile and all rows lying within $2\sigma$ of the centroid of the Gaussian fit were averaged together to create the final spectrum. This corresponds to those rows lying within 1.2 $R_J$ of the centrifugal equator. Spectra created in this fashion were averaged together to produce the spectrum in Figure 3.

**[FIGURE 3]**

This spectrum, which contains data from both EUV and FUV channels, covers a wavelength range of 561Å-1912Å. It is the average of 17 individual 1000-second images covering a range of projected radial distances from 4-8 $R_J$. The major spectral features are labeled by approximate wavelength and the ion species responsible for the majority of the emission in each feature. Below the spectrum are plotted the wavelengths of the radiative transitions produced by the five major ion species of the Io torus: O II, S III, S II, SIV, and O III, as contained in the CHIANTI v. 4.2 atomic physics database (Dere *et al.* 1997, Young *et al.* 2003). The wavelength region covered by UVIS contains over 500 individual radiative transitions from these five ion species. The high density of emission lines, coupled with the ~3Å spectral resolution of UVIS, means that with only three exceptions—the S IV line at 1063Å and the S III multiplet at 1713Å and 1729Å—all the features observed in the UVIS spectra are blends of the multiplet structure within a particular ion species, blends of radiative transitions from two or more different ion species, or some combination thereof.

This spectral complexity necessitates a detailed, multi-species model to properly interpret the data.

**III. TORUS SPECTRAL EMISSIONS MODEL**

In order to model the torus spectra, we developed a homogeneous, 0-D "cubic centimeter" spectral emission model. Such a model calculates the volume emission rate for a given spectral line, i.e. the number of photons at a specific wavelength produced by a single cubic centimeter of plasma in one second, and integrates this over the line of sight to produce a synthetic spectrum. The technique is similar to that used by Shemansky (1980) and Shemansky and Smith (1981). The brightness, B of a given spectral line is given by:

(1) $\quad B = 10^{-6} \int A_{ji} f_j(T_e, n_e) n_{ion} dl \quad$ Rayleighs

where $A_{ji}$ is the Einstein coefficient for spontaneous emission, $f_j$ is the fraction of ions in state j, $T_e$ is the electron temperature, $n_e$ is the electron number density, $n_{ion}$ is the number density of the ion species responsible for the emission, and the integral is over the line of sight. The level populations, $f_j$, are determined by solving the level balance equations for each ion species in matrix form:

(2) $\quad$ **Cf = b**

where f is a vector containing the fraction of ions in a particular energy state, relative to the ground state; b is a vector whose elements are all zero except for the first element, which is equal to one; and C is a matrix containing the rates for collisional excitation and de-excitation and radiative de-excitation. The elements of this matrix are given by:

(3) $\quad C[i, j] = A_{ij} + n_e q_{ij}$

where $A_{ij}$ is the Einstein coefficient for spontaneous emission if state i is at a higher energy than state j and zero otherwise. $q_{ij}$ is the rate coefficient for collisional excitation (or de-excitation) from state i to state j and is given by:

$$(4) \quad q_{ij} = \int_0^\infty \hat{g}_e v \sigma_{ij} \, dv$$

$\hat{g}_e$ is the normalized distribution function, v is the electron velocity, and $\sigma_{ij}$ is the cross-section for the transition from state i to state j. Once Eq. 2 has been solved, the level populations vector, f, is re-normalized so that the sum of its elements is equal to one.

### A. Thermal and Non-Thermal Electron Distributions

If the electrons are distributed according to Maxwell-Boltzmann statistics:

$$(5) \quad \hat{g}_e(v) = 4\pi^{-1/2} \left( m_e / 2kT_e \right)^{3/2} v^2 \exp\left(-m_e v^2 / 2kT_e\right)$$

where $m_e$ is the mass of an electron and k is the Boltzmann constant. With this equation for the electron distribution function, Eq 4. reduces to:

$$(6) \quad q_{ij} = 2\pi^{1/2} a_0 \hbar m_e^{-1} w_i^{-1} \left( I_\infty / kT_e \right)^{1/2} \Upsilon_{ij} \exp\left( E_{ij} / kT_e \right)$$

where $2\pi^{1/2} a_0 \hbar m_e^{-1} = 2.1716 \times 10^{-8}$ cm$^3$ s$^{-1}$; $w_i$ is the statistical weight of state i; $I_\infty = 13.6086$ eV; and $E_{ij}$ is the transition energy between states i and j. $\Upsilon_{ij}$ is the thermally-averaged collision strength, as defined by Seaton (1953) and is given by:

$$(7) \quad \Upsilon_{ij} = \int_0^\infty \Omega_{ij} \exp\left(-E_j / kT_e\right) d\left(E_j / kT_e\right)$$

where $E_j$ is the electron energy after the collision and $\Omega_{ij}$ is the collision strength, which is related to the collision cross-section, $\sigma_{ij}$, by:

$$(8) \quad \sigma_{ij} = \pi \hbar^2 \Omega_{ij} / m_e^2 v^2 w_i$$

In the basic form of our spectral emissions model, the electron distribution is Maxwellian and therefore defined by a single parameter, $T_e$. However, in-situ measurements of the electron distribution in the Io plasma torus made by the Voyager and Galileo spacecrafts suggest that the electron distribution function in the Io torus may actually be non-thermal or at least have a non-thermal, high-energy tail (Sittler and Strobel 1987, Frank and Paterson 2000). Rather than examining the changes in the torus spectrum due to an arbitrary, non-thermal electron distribution, we have focused our efforts on modeling the effects of a kappa electron distribution function (Vasyliunas 1968). Kappa distributions have been invoked to explain discrepancies between spectra from the Voyager Ultraviolet Spectrometer (UVS) and model spectra based on emission rates generated by the Collisional and Radiative Equilibrium code (COREQ) (Taylor 1996), differences in the in-situ plasma measurements made by the Voyager and Ulysses spacecrafts (Meyer-Vernet *et al.*1995, Moncuquet *et al.* 2002), latitudinal changes in torus ion temperature in ground-based observations of the torus (Thomas and Lichtenberg 1997), and features of Io's ultraviolet limb-glow (Retherford *et al.* 2003). Kappa distributions are also attractive because a physical mechanism has been proposed to produce them in space plasmas (Collier 1993). A review of kappa distributions and their effect on astrophysical plasmas is given by Meyer-Vernet 2001.

A kappa distribution, defined by the equation:

(9) $\quad \hat{g}_e(v) = 4\pi^{-1/2} \left(m_e / 2\kappa k T_e\right)^{3/2} \left(\Gamma(\kappa+1)/\Gamma(\kappa-\tfrac{1}{2})\right) v^2 \left(1 + mv^2 / 2\kappa k T_e\right)^{-(\kappa+1)}$

is quasi-Maxwellian at low temperatures but falls off as a power law at high temperatures. From a computational perspective, a kappa distribution has an additional advantage over other types of non-thermal distributions in that it is fully defined by only two parameters: the characteristic temperature of the distribution, $T_c$, and the parameter, $\kappa$. The characteristic temperature, $T_c$, of a kappa distribution is related to the energy at the peak of the distribution function. Unlike a Maxwellian distribution, the characteristic temperature in a kappa distribution is not the same as the effective temperature, $T_e$, which is related to the mean energy per particle of the distribution. Instead, $T_c$ and $T_e$ are related by the equation:

(10) $\quad T_e = T_c \kappa/(\kappa-3/2)$

As can be seen from Eq. 10, the κ-parameter determines the degree to which the distribution is non-Maxwellian. The larger the value of the kappa parameter, the closer the distribution is to a Maxwellian and in the limit of $\kappa = \infty$, the distribution is equivalent to a Maxwellian.

The atomic data required for these calculations (wavelengths, energy levels, A coefficients, thermally-averaged collision strengths, etc.) are obtained from the CHIANTI database (Dere *et al.* 1997, Young *et al.* 2003). CHIANTI consists of a set of critically evaluated atomic data together with a set of routines written in the Interactive Data Language (IDL) to calculate emission spectra from astrophysical plasmas. The database is a compilation of both experimental and theoretical values and is periodically updated. Version 4.2 of CHIANTI was used for all modeling in this paper. The CHIANTI database implicitly assumes a Maxwellian distribution and contains only thermally-averaged collision strengths, $Y_{ij}$, stored according to the method of Burgess and Tully 1992. The cross sections, $\sigma_{ij}$, are required to evaluate Eq. 4 if the distribution function, $\hat{g}_e$, is non-Maxwellian. When using a kappa distribution, we must therefore approximate the integral in Eq. 4 as the linear combination of five thermally-averaged rate coefficients, $q_{ij}$:

(11) $\quad q'_{ij} = \sum_{k=1}^{5} w_k q_{ij}(T_k)$

where $q_{ij}(T_k)$ is the rate coefficient, given by Eq. 6 with $T_e=T_k$ and $w_k$ is the relative weighting of the rate coefficient. The weights, $w_k$, are determined by logarithmically fitting the linear combination of five Maxwellians to a kappa distribution over the energy range of 0.01-500 eV. The resulting fit is within 10% of the value of the kappa distribution over the entire energy range. It is worth reiterating that when we fit the spectra, we solve only for two parameters, $T_C$ and $\kappa$, that fully describe the kappa distribution; the $w_k$'s and $T_k$'s in Eq. 11 are completely determined by the values of $T_C$ and $\kappa$.

The decision to use the CHIANTI database over other means of determining radiative emission rates, namely the Collisional and Radiative Equilibrium (COREQ) code that is an extension of the work of Shemansky and Smith (1981), was made based on the public availability, documentation, periodic updating, and ease of use of the CHIANTI database. For most spectral features in the UVIS wavelength range, the differences between models using CHIANTI and models using COREQ are at the 10% level, with COREQ generally predicting more emission than CHIANTI (D.E. Shemansky, personal communication). However, for several spectral features (e.g. S III 1021Å, S IV 1063Å, and S II 1260Å) there exist large (factors of several) differences between the emissions predicted by the two databases. One notable weakness of the CHIANTI database—at least as it exists in version 4.2—is that it does not include radiative transitions from singly ionized sulfur (S II) at wavelengths less than 765Å. As a result, the S II features at 642Å and 700Å are absent from our model.

**B. Line of sight Assumptions**

Evaluating the integral in Eq. 1 requires knowledge of how $T_e$, $n_{ion}$, and $n_e$ vary over the line of sight. Since these are the very quantities we are trying to derive from the spectra, certain assumptions must be made. As indicated by Eq. 1, the level populations of the ions, $f_j$, are a function of the electron density. In theory, this dependence can be used as a diagnostic of the local torus electron density (Feldman *et al*, 2004). In practice, however, the spectral resolution of UVIS is insufficient to resolve the density-sensitive multiplet structure present in the torus spectra, and therefore, the torus spectra observed by UVIS are effectively independent of the local electron density.

We have chosen to use a relatively simple treatment of projection effects. We make the assumption that the electron distribution function of the torus, be it Maxwellian or kappa, is uniform over the line of sight. This assumption should not significantly affect our results for two reasons. First, the observed brightness of the torus falls off sharply with radial distance outside of 6 $R_J$ (Brown, 1994; paper I). Second, because we are observing the torus at its ansa, the pathlength of the line of sight through regions of the torus lying exterior to the region of interest is minimized. The combination of lower brightness and smaller pathlengths mean that the spectral contributions from regions of the torus lying exterior to the region we are interested in will be relatively small. Inside of 6 $R_J$, however, the local electron temperature is too low to excite much emission in the EUV/FUV. In this region, line of sight projection effects become much more important as the majority of observed EUV/FUV photons are actually emitted from regions of the torus lying at greater radial distances than the ansa, and the validity of our assumption breaks down. Therefore, we have limited our analysis to those regions lying outside of 6 $R_J$.

With the assumption of a uniform electron distribution over the line of sight, Eq. 1 reduces to:

(12) $\quad B = 10^{-6} A_{ji} f_j(T_e) N_{ion} \quad$ Rayleighs

where $N_{ion} = \int n_{ion} \, dl$ is the ion column density. The ion column densities, $N_i$, needed to match the observed torus brightness depend on the level populations of the ion species, $f_j$, which, in turn, depend on the shape of the electron distribution function.

The plasma composition of our model is specified by six parameters, one for the column density of each of six ion species: S II, S III, S IV, S V, O II, and O III, For computational reasons, as well as to reduce correlations between parameters, we have found it advantageous use five parameters for the ion column densities relative to the column density of S III ($N_{S\,II}/N_{S\,III}$, $N_{S\,IV}/N_{S\,III}$, $N_{S\,V}/N_{S\,III}$, $N_{O\,II}/N_{S\,III}$, $N_{O\,III}/N_{S\,III}$) and a sixth parameter for the electron column density, $N_e$. The column density of S III is then derived from the charge neutrality condition:

$$(13) \quad \sum_{ions} q_{ion} N_{ion} = N_e$$

where $q_{ion}$ is the charge on each ion. Protons are included in the calculation of charge neutrality at the 0.1 $N_e$ level (Bagenal 1994). In addition to the six parameters for the plasma composition, the model requires one parameter ($T_e$) to specify the electron distribution if we are using a thermal distribution, or two parameters ($T_c$ and $\kappa$) for a kappa distribution.

With these parameters and the above equations we produce model spectra and fit them to the data by minimizing the $\chi^2$ statistic using a combination of Levenberg-Marquardt least squares and downhill simplex (amoeba) algorithms (Press *et al.* 1992). This combined approach was necessary because the Levenberg-Marquardt method, while computationally efficient, tended to get stuck in small, local $\chi^2$ minima of the seven-dimensional parameter space. The downhill simplex method was more successful at finding the global minimum value for $\chi^2$, at the expense of greatly increased computation time. Therefore, we began the fitting procedure using the Levenberg-Marquardt algorithm. Once the algorithm had settled in to a local minimum in parameter space we used the fit parameters to specify one point in the initial input simplex. The remaining points of the simplex consisted of the Levenberg-Marquardt algorithm fit parameters plus a random deviation. With these inputs, the downhill simplex algorithm was generally able to climb out of local minima in parameter space and find a lower overall value of the $\chi^2$ statistic.

## IV. RESULTS

A typical fit of the spectral model to an individual UVIS spectrum of the Io torus can be seen in Fig. 4.

**[FIGURE 4]**

This spectrum of the ansa at $6.3R_J$ is typical of the quality of the data and quality of the model fit There is generally good agreement between the model and the data. However, the S IV 657Å, S III 702Å, S II 910Å, and S III 1729Å features are consistently underfit by the model. The discrepancies between the spectral emissions model and the UVIS spectra are likely caused, at least in part, by inaccuracies in the atomic data contained in the CHIANTI database. Herbert *et al.* (2001) report similar discrepancies in their analysis of EUVE spectra of the Io plasma torus.

Preliminary work on the in-flight calibration of the EUV channel of UVIS suggests that the true instrumental effective area below 740Å may be greater than what was measured in the laboratory by as much as a factor of two (D.E. Shemansky and W.E. McClintock, personal communication). If the true instrumental effective area below 740Å has been underestimated than the brightness of the spectral features in this region has been overestimated, which would bring the features at 657Å and 702Å into better agreement with the model. However, given the relatively good fit of the S III features at 680Å and 729Å, the shape of the instrumental effective area curve (see paper I) would have to change dramatically to fully reconcile the differences between model and spectra at 657Å and 702Å while preserving the quality of the fit elsewhere. It therefore seems most likely that these discrepancies are primarily caused by inaccuracies in the atomic physics data.

### A. Electron Temperature and Densities

The electron temperature, $T_e$, and column density, $N_e$, derived from the Cassini UVIS torus spectra are shown in Fig. 5.

**[FIGURE 5]**

For comparison to the Voyager era, we also plot these quantities as obtained by the model of Bagenal (1994), hereafter referred to as B94, which is based on an analysis of Voyager 1 Plasma Science (PLS) data coupled with the ion composition derived from analysis of the Voyager 1 UVS spectra by D.E. Shemansky. Independent analysis of Voyager 1 UVS spectra of the torus was conducted by Herbert and Sandel (2000), which is hereafter referred to as HS00. The electron temperatures derived from the UVIS spectra are somewhat lower than those from the models of B94 and HS00. Although it is not plotted, HS00 generally has a slightly higher electron temperature than B94. The sharp increase in electron temperature between 7.4 $R_J$ and 8.5 $R_J$ present in the B94 model is not seen in the UVIS spectra, nor was it seen by HS00. Our results support the claim made by HS00 that this sudden increase in electron temperature is not representative of "typical" conditions in the Io torus. Curiously, the UVIS electron temperature profile reaches a minimum value of 4.43 eV at 6.6 $R_J$. A similar dip is seen in the electron temperature profile of HS00, but not in B94.

The derived electron column density, $N_e$, is plotted in the lower panel of Fig. 5. The column density falls off monotonically with increasing radial distance from a maximum value of just under $10^{14}$ electrons cm$^{-2}$ to $5 \times 10^{12}$ electrons cm$^{-2}$ near 9 $R_J$. Inside of 7.5 $R_J$, the UVIS values generally lie within one error bar of the values derived from B94. Outside of 7.7 $R_J$ the column density falls off more rapidly with distance than the B94 model.

Although the electron column density is the quantity that is actually measured by remote sensing instruments, often we would like to know the local electron number density. In order to extract this information from the integrated column density, we must make additional assumptions about how the torus plasma is distributed along the line of sight. The first additional assumption we make is that the relative plasma composition is uniform over the line of sight (previously we had assumed only that the electron distribution function was uniform over the line of sight). We then assume that the local electron density as a function of radial distance is reasonably well described as a power law:

$$(14) \quad n_e(r) = \begin{cases} \alpha_6 (r/6)^{-\beta_1} & \text{for } r \leq r' \\ \alpha_8 (r/8)^{-\beta_2} & \text{for } r > r' \end{cases}$$

where r is the radial distance, measured in $R_J$. We then varied the parameters $\alpha_6$, $\beta_1$, $\alpha_8$, $\beta_2$, and r' to fit the integral of $n_e(r)$ over the line of sight to the derived values of the electron column density, subject to the constraint that $n_e(r)$ be continuous at r = r'. The resulting fit to the integrated column density is show in Fig. 6.

**[FIGURE 6]**

Also shown are the function $n_e(r)$ and the electron density profile of B94. Although the value of the curve itself is slightly less than the B94 value, between 6 $R_J$ and 7.4 $R_J$, the slope of the electron density curve derived from the UVIS spectra is almost identical to the slope of the B94 model. The UVIS electron density is less than the electron density derived by HS00 and a factor of ~2 less than the electron density derived by the Galileo Plasma Wave Subsystem during the J0 flyby (Gurnett *et al.*, 1996).

We derive the number of electrons per shell of magnetic flux using the equation:

(15) $\quad NL^2 = 4\pi R_J^3 L^4 \int_0^{\theta_{max}} n_e(\theta) \cos^7(\theta) d\theta \approx 2\pi^{3/2} R_J^3 n_e(\theta=0) H L^3$

where L is the radial distance of a magnetic field line at the magnetic equator, $\theta$ is the magnetic latitude, $n_e(\theta=0)$ is given by Eq. 14, and H is the scale height given by:

(16) $\quad H = \left(2k\bar{T}_{ion}(1+\bar{Z}_{ion}T_e/\bar{T}_{ion})/3m\Omega_J^2\right)^{1/2} = 0.64\left(\bar{T}_{ion}(1+\bar{Z}_{ion}T_e/\bar{T}_{ion})/\bar{A}_{ion}\right)^{1/2} R_J$

where $\bar{T}_{ion}$ is the average ion temperature, $\bar{Z}_{ion}$ is the average charge per ion, and $\bar{A}_{ion}$ is the average ion mass number. The average ion temperature, $\bar{T}_{ion}$, cannot be directly determined from our analysis of the UVIS spectra so we use the values from B94 (60 eV from 6.0-7.5$R_J$, and increasing roughly linearly from 7.5 $R_J$ to a value of 228 eV at 9.0 $R_J$). Since the scale height, H, varies as $\bar{T}_{ion}^{1/2}$, this assumption should not significantly affect our calculation of flux tube content. The derived values for $NL^2$ as a function of radial distance are fit well by a single power law: $NL^2(r)=2.0\times10^{36}(r/6)^{-2.1}$. The index of the power law for the UVIS-derived value of flux tube content, 2.1±0.4, is statistically identical to the value derived by B94, and significantly less than the value of 3.5 derived by Herbert and Sandel (1995). An index of 2 is consistent with flux tube interchange as the mechanism for radial transport of plasma (see review by Thomas *et al.*, 2004 and references therein). There is some evidence to suggest that the index of the power law fit to the UVIS-derived flux tube content, $NL^2$, is greater than two outside of 7.5 $R_J$. However, this finding is only marginally statistically significant.

**B. Ion Mixing Ratios.**

The torus composition derived from the UVIS spectra obtained from the January 14, 2001 radial scan is plotted in Fig 7.

**[FIGURE 7]**

We have plotted the derived composition information as ion mixing ratios, (i.e. ion densities divided by the electron density). For comparison to the Voyager era, we have also plotted the mixing ratios from B94. The UVIS-derived composition is significantly different than the Voyager values, implying a fundamental change in torus composition between the two epochs. This is hardly surprising, given that substantial compositional changes were observed during the six months of the Cassini Jupiter flyby (see paper I). It is important, then, to remember that the compositional information presented in this paper comes from observations made during single day, January 14, 2001.

The torus observed by UVIS contains substantially less oxygen than the torus of the Voyager epoch. The total $O_i/S_i$ ion ratio, averaged between 6 $R_J$ and 8 $R_J$, is 0.9, compared to 1.6 in B94. The sharp decrease in the amount of oxygen in the torus relative to the Voyager 1 conditions supports the findings of ground-based optical observations of the Io torus (Morgan, 1985; Thomas *et al.*, 2001) and is opposite to the higher oxygen levels found by Galileo PLS on the J0 flyby (Crary *et al.*, 1997) and EUVE (Herbert *et al.*, 2001). The UVIS composition shows a trend toward higher ionization states: the mixing ratios of S II and O II derived from the UVIS spectra are both lower than the B94 values, while the mixing ratios of S III, S IV, and O III are generally higher. This results in an increase in the average charge per torus ion, $<Z_i>$, to 1.7 compared with a value of 1.4 in the B94 model.

Determination of the relative ion abundance of O II and O III from EUV spectra has been historically difficult (Brown *et al.* 1983). This is due primarily to the paucity of bright emission lines from these ions in the EUV/FUV region of the spectrum. In marked contrast to the sulfur ion species present in the torus, O III has just three relatively bright spectral features in the wavelength range covered by UVIS: the brightest centered at 834Å and the other two at 703Å and 1666Å. Singly ionized oxygen has but one bright spectral feature, located at 833Å. Initial analysis of the Voyager UVS spectra focused on determining the abundance of O III by fitting to the multiplet at 703Å. The O II abundance was then derived by determining the extra emission required to fit the feature at 833Å. Unfortunately, the O III multiplet at 703Å is heavily blended with significantly brighter emissions from S III centered on 702Å. Thus, this approach requires knowledge of the amount of S III along the line of sight and accurate atomic data for O II, O III, and S III. These difficulties led to the initial analyses of Voyager UVS spectra concluding that the ratio of O II to O III in the Io torus was less than 1 (Shemansky 1980, Shemansky and Smith 1981, Broadfoot *et al.* 1981). Since that time, numerous additional analyses of torus observations at UV and optical wavelengths have confirmed that O II is actually the dominant ionization state of oxygen, with O III being a relatively minor constituent (Brown *et al.* 1983, Smith and Strobel 1985, Shemansky 1987, McGrath *et al.* 1993, Thomas 1993, Hall *et al.* 1994, Herbert *et al.* 2001, and others).

If we consider only the EUV channel of UVIS, the spectral emissions model concludes that O III is the dominant ionization state of oxygen in the Io torus. This unphysical result occurs because the model maximizes the amount of O III in order to minimize the model/spectrum discrepancy at 702Å (see Fig. 4). With the inclusion of the FUV channel, there are two additional O III spectral lines located at 1661Å and 1666Å. These lines, first detected in the Io torus by Moos *et al.* (1991), place a strong constraint on the amount of O III present in the torus. Unfortunately, they are relatively faint and barely above the level of noise in the UVIS spectra. Therefore, the values we derive for the mixing ratio of O III (O II) as a function of radial distance should more properly be thought of as an upper (lower) limit on the actual value. With this caveat in mind, there is still significantly more O III and less O II compared to the Voyager model of B94. The [O II] / [O III] ratio, averaged over 6.2-8.8 $R_J$, is 3.7—less than half the corresponding value of 8.8 from B94. The value of this ratio generally decreases with increasing radial distance, which is consistent with the observed increase in electron temperature. The upper limit on the amount of O III seen in the UVIS spectra is still significantly less than the lower limit reported by Crary *et al*. (1998) during the Galileo spacecraft's flythrough of the Io torus in 1995.

In the EUV/FUV region of the spectrum, the brightest emission feature (by over two orders of magnitude) due to S V, occurs at 786Å. Since the 786Å S V feature lies between several nearby spectral features from S II and S III it has proven difficult to detect. The initial analysis of Voyager UVS spectra of the Io torus placed an upper limit of 11 $cm^{-3}$ on the mean ion number density of S V (Shemansky and Smith 1981). The factor-of-ten increase in spectral resolution of the Cassini UVIS over the Voyager UVS us to make what we believe to be the first spectroscopic detection of S V in the Io torus. Near 6 $R_J$, where the signal-to-noise ratio is highest, S V is detected at the 3-$\sigma$ level. S V is a trace component of the torus, present at a mixing ratio of 0.003 at 6 $R_J$ and rising to maximum of 0.01 at 8.5 $R_J$. Another instrument aboard the Cassini spacecraft, the Charge-Energy-Mass Spectrometer (CHEMS) of the Magnetospheric Imaging Instrument (MIMI), detected S V ions on January 10 and January 23, 2001—periods when the spacecraft was within the magnetosphere of Jupiter (Hamilton *et al*. 2001, Krimigis *et al*. 2001). While the MIMI result does not directly confirm the detection of S V ions in the Io torus, it does confirm that S V is present within the Jovian magnetosphere.

### C. Uncertainties in Derived Model Parameters

The error bars presented in Figs. 6 and 7 represent the formal 1-$\sigma$ error bars of the least-squares fit, i.e. they are the square roots of the diagonal elements of the covariance matrix. This method of estimating errors implicitly assumes that the model parameters are independent of each other. However, many of the model parameters are correlated (or anti-correlated) e.g. electron column density and electron temperature. In order to assess the effect of parameter correlations on the actual uncertainty in the model parameters, a series of two-dimensional confidence intervals was generated following the method of Press *et al.* (1992). Four of these confidence intervals for the spectrum at 6.2 $R_J$ can be found in Fig. 8.

**[FIGURE 8]**

The cross in the center of the $\Delta\chi^2$ contours represents the size of the formal error bar. The top two panels, $N_{S\,IV}/N_{S\,III}$ vs. $N_{S\,II}/N_{S\,III}$ and $N_e$ vs. $N_{S\,IV}/N_{S\,III}$, provide examples of parameters that are minimally correlated, while the bottom two panels show pairs of parameters that are strongly anti-correlated. The 1-D confidence interval for a single parameter is defined by the projection of the contour of desired probability onto that parameter's axis. For example, the probability that the "true" value of $N_{O\,III}/N_{S\,III}$ lies in the interval 0.065-0.255 is 68%. The formal error bars almost always underestimate the full extent of the parameter confidence intervals, so the error bars in Figs. 6 and 7 should be used with some caution.

### D. κ-Distribution Results.

As described above, the spectral emissions model used to fit the UVIS spectra can accommodate either a thermal, Maxwellian electron distribution function or an approximation to a non-thermal, kappa electron distribution function. Fits of the spectra were made using both distribution functions. The models that used a Maxwellian distribution and the models that used a kappa distribution both produced fits to the data qualitatively similar to Fig. 4. However, the value of the $\chi^2$ statistic was marginally lower (~2%) for the models using a kappa distribution, indicating a somewhat better fit. The torus ion composition derived by the two models was statistically identical—a surprising result. It appears that the derived ion mixing ratios are nearly independent of the shape of the electron distribution function for most "reasonable" distribution functions. This effect can also be seen in the relatively large error bars for the electron parameters derived from the kappa distribution model.

Although the ion composition between the two models was indistinguishable, the models using the kappa approximation required an electron column density ~1.7 times greater than the models that used a Maxwellian to fit the spectra. The reason for this can be understood by examining the shape of the distribution functions. Figure 9 shows the two best-fit distribution functions for the spectrum of the torus obtained at 7.4 $R_J$.

**[FIGURE 9]**

From Eq. 12 we see that the observed brightness of the torus spectrum is dependent on the level populations of the ions, which from Eqs. 2 and 3 will depend on the shape of the electron distribution function. For the example shown in Fig. 9, the kappa distribution function is greater than the Maxwellian distribution function below 5 eV and above 60 eV. However, electrons with energies of 5 eV or less are generally incapable of collisionally exciting ions to the states that produce EUV/FUV photons. As a result these electrons have little effect on the observed EUV/FUV spectrum. Electrons in the high-energy tail of the kappa distribution are certainly capable of exciting EUV/FUV transitions, but there are far fewer of these electrons than there are electrons in the 5-60 eV range. In this critical middle energy range, the Maxwellian distribution has more electrons than the kappa distribution. As a result, the kappa distribution model requires higher ion column densities than the Maxwellian distribution model in order to match the observed brightness of the spectrum. The similar $\chi^2$ statistic of models using the two different distribution functions (Maxwellian and kappa) implies that the shape of the electron distribution can not be tightly constrained by EUV/FUV observations of the torus alone. The electron distribution function could be better constrained by either obtaining an independent measure of the ion column densities or extending the wavelength range of the analysis into the optical.

In February 1992, the Ulysses spacecraft flew through the Io torus. This pass through the torus is unique in that the spacecraft trajectory was basically north-to-south, as opposed to lying close to the equatorial plane. For the period when ULYSSES was within 15° of the Jovian equator, it sampled the region from approximately 7.1-8.2 $R_J$ in radial distance. Although the particle detector instruments were not turned on for this encounter, in situ measurements of the electron density and temperature were made by the Unified Radio and Plasma (URAP) wave experiment (Stone *et al.* 1992a, Stone *et al.* 1992b). Analysis of this data revealed that the bulk electron temperature was not constant along magnetic field lines, but rather varied with latitude in anticorrelation with density (Meyer-Vernet *et al.*, 1995; Moncuquet *et al.*, 2002). The authors proposed that this effect could be explained if the electron distribution approximated a kappa distribution with $\kappa= 2.4 \pm 0.2$.

The values for $\kappa$ derived from the UVIS spectra are shown in Fig. 10.

[FIGURE 10]

Outside 6.6 $R_J$, the values for κ show a steady decrease with radial distance. The Ulysses URAP value of κ= 2.4 ± 0.2, which was measured at ~8$R_J$, fits nicely between the UVIS values derived at 7.9 and 8.1 $R_J$. The decrease of kappa inside 6.6 $R_J$ may result from a projection of the outer regions of the torus into the line of sight.

## V. CONCLUSIONS

We have analyzed a radial scan of the midnight sector of the Io plasma torus obtained by the Ultraviolet Imaging spectrograph on January 14, 2001. These observations record the radial structure of Io torus at a local time of 01:50, which has not been previously observed. Two dimensional spectrally dispersed images of the torus are obtained from the UVIS instrument, although to increase the signal-to-noise, we average over the latitudinal structure of the torus. Features from six different ion species are readily apparent in the torus spectra.

In order to derive information about the plasma composition from the spectra, we developed a spectral emissions model, similar to that used by Shemansky and Smith (1981), which incorporates the latest atomic physics data from the CHIANTI database (Dere *et al.*, 1997; Young *et al.*, 2003). In order to deal with line of sight projection effects, we assume that the electron distribution function is uniform over the column through the torus, an assumption that should not significantly affect our results. We find that the electron temperature is less than that predicted by the Voyager era model of Bagenal (1994). We find that the observed radial profile of electron column density is well matched by assuming that the local electron number density profile is proportional to $r^{-5.4}$ from 6.0-7.8 $R_J$ and $r^{-12}$ outside of 7.8 $R_J$. If we use this profile for electron density and the ion temperatures derived by Bagenal (1994) we find that the flux tube content of the Io torus is proportional to $r^{-2}$, which is consistent with flux tube interchange acting to transport plasma radially outward.

The plasma composition derived from the UVIS spectra of January 14, 2001 is significantly different that the torus composition during the Voyager era. However, paper I has shown significant temporal variations over the six-month flyby of Jupiter. Both O II and S II are depleted compared to the Voyager values, while S III and S IV show enhancements. The O/S ion ratio of 0.9, obtained from the UVIS spectra, is much lower than the Voyager value of 1.6. Ground-based observations of the torus have also found less oxygen than predicted by the Voyager models. In addition to the lower O/S ratio, we find that the charge per ion has increased to 1.7 from 1.4. The spectral resolution of UVIS allows us to report the $3\sigma$ detection of S V. S V, which has not previously been detected in the Io torus, is present in the torus at a mixing level of ~0.5%.

Our spectral emissions model has the ability to approximate the effects of an arbitrary, non-thermal electron distribution as the linear combination of Maxwellian components. We explored the effects of using a non-thermal kappa distribution, which is quasi-Maxwellian at low energies and a power law at high energies, to analyze the torus spectra. Models using a kappa distribution of electrons had a marginally lower value of the $\chi^2$ statistic, although the actual spectral fits were qualitatively very similar to those produced by the Maxwellian model. We found that the ion composition derived using the kappa distribution model was identical to the ion composition derived using a Maxwellian model. However, as a result of the shape of the distribution function in the 5-60 eV range of energy, the kappa models required a higher electron column density to match the brightness of the UVIS torus spectra. The value of the $\kappa$ parameter, which determines the index of the power law, high-energy tail of the distribution, was found to generally decrease with radial distance. The derived radial profile value of the κ parameter is consistent with the measurement of κ=2.4 at 8 $R_J$ made by the Ulysses URAP instrument (Meyer-Vernet *et al.*, 1995).

The analysis presented this data set has focused on the radial variations of torus parameters. However, the orientation of the UVIS entrance slits parallel to the Jovian rotational axis also make these data well suited to analyze the latitudinal structure of the torus. Such a latitudinal analysis will be the focus of future work.

**ACKNOWLEDGEMENTS**

Analysis of the Cassini UVIS data is supported under contract JPL 961196. FB acknowledges support as Galileo IDS under contract JPL959550. The authors wish to thank Bill McClintock and the rest of the UVIS science and operations team for their support. We thank Craig Markwardt for use of the MPFITFUN IDL routine and Peter Young for his help with the CHIANTI database. CHIANTI is a collaborative project involving the NRL (USA), RAL (UK), and the Universities of Florence (Italy) and Cambridge (UK).

Krimigis, S.M., and 18 colleagues 2001. Observations in Jupiter's vicinity with the Magnetospheric Imaging Instrument (MIMI) during Cassini/Huygens flyby (October 2000-March 2001). paper presented at AGU Spring Meeting 2001, Boston, 2001.

Lichtenberg, G. 2001, *Massenbilanz und Energiehaushalt des Io-Plasma-Torus: Modell und Beobachtung*, Ph.D. Thesis, University of Göttingen, Göttingen, Germany.

Lichtenberg, G., N. Thomas, and T. Fouchet 2001. Detection of S (IV) 10.51 μm emission from the Io plasma torus. *J. Geophys. Res.* **106**, 26899-29910.

McClintock, W.E. G.M. Lawrence, R. A. Kohnert, and L.W. Esposito 1993. Optical design of the Ultraviolet Imaging Spectrograph for the Cassini mission to Saturn. *Opt. Eng.* **32**, 3038-3046.

McGrath, M.A., P.D. Feldman, D.F. Strobel, H.W. Moos, and G.E. Ballester 1993. Detection of [O II] λ2471 from the Io plasma torus. *Astrophys. J.* **415**, L55-L58.

Meyer-Vernet, N, M Moncuquet, and S. Hoang 1995. Temperature inversion in the Io plasma torus. *Icarus* **116**, 202-213.

Meyer-Vernet, N. M. 2001. Large scale structure of planetary environments: the importance of not being Maxwellian. *Planet. Space Sci.,* **49**, 247-260.

**Table 1.** Observational Parameters

| Date | Time (U.T.) | $R^a$ | $\lambda_{III}{}^b$ |
|---|---|---|---|
| 14-JAN-2001 | 16:31:04 | 8.84 | 252 |
| 14-JAN-2001 | 17:04:24 | 8.51 | 272 |
| 14-JAN-2001 | 17:37:44 | 8.14 | 292 |
| 14-JAN-2001 | 18:02:44 | 7.85 | 308 |
| 14-JAN-2001 | 18:19:24 | 7.66 | 318 |
| 14-JAN-2001 | 18:36:04 | 7.46 | 328 |
| 14-JAN-2001 | 18:52:44 | 7.25 | 338 |
| 14-JAN-2001 | 19:09:24 | 7.03 | 348 |
| 14-JAN-2001 | 19:26:04 | 6.82 | 358 |
| 14-JAN-2001 | 19:42:44 | 6.60 | 8 |
| 14-JAN-2001 | 19:59:24 | 6.38 | 18 |
| 14-JAN-2001 | 20:16:04 | 6.16 | 28 |

[a] Projected radial distance to center of UVIS entrance slit in $R_J$
[b] System III longitude of torus ansa

**FIGURE CAPTIONS**

**Figure 1.** Observing geometry for the UVIS observation of the Io torus on 14-January-2000. The initial and final positions of the projected field of view of the UVIS EUV channel entrance slit, Io, and Europa are shown. Jovian north is up. The local solar time of the spacecraft is 19:40. The Cassini spacecraft is south of the Jovian equator, so the far side of the obits appear below the near side..

**Figure 2.** Spectral image of the Io torus at 6.5 RJ. The EUV channel appears above the FUV channel. For the observations presented in this paper, the long axis of the slit was oriented roughly perpendicular to Jupiter's equator. North is up and Jupiter is to the left. The data have been background-subtracted, flatfielded, and calibrated to physical units. The region from 1210-1230Å in the FUV channel is dominated by Lyman-α from the interplanetary medium and has been set to zero. The spatial scale is 0.24 $R_J$/pixel in the vertical (spatial) direction and 0.06 $R_J$/pixel in the horizontal (spectral) direction.

**Figure 3.** Composite spectrum of the Io plasma torus from 561Å-1913Å. 1-D spectra of the Io torus were created by averaging the rows of the 2-D spectral images lying within 1.2 RJ of the latitudinal center of the torus. The composite spectrum was created by averaging together 17 individual 1-D spectra of the torus covering a radial range of 4-8 RJ. The spectral features are labeled and color-coded by the ion species that makes the dominant contribution to the feature. Locations (as contained in the CHIANTI database) of the individual spectral lines of the five major ion species in the torus are plotted beneath the spectrum.

**Figure 4**. Sample fit of the model to a UVIS spectrum of the Io torus at 6.3 $R_J$. The model generally fits well to the spectrum with the exception of the three features at 657Å, 702Å, and 729Å, which are consistently underfit. The region from 1210-1230Å is dominated by Lyman-alpha emission from the interplanetary medium and has been set to zero.

**Figure 5.** Best-fit electron parameters as a function of radial distance. The electrons distribution is assumed to be a single Maxwellian. The solid lines are the UVIS results, while the dotted lines are the parameters from the Voyager-based model of Bagenal (1994). The error bars are the formal 1-σ errors obtained from the fitting algorithm. The electron column density is derived from the ion column densities using the charge-neutrality condition

**Figure 6.** The derived electron column density (points with error bars) plotted versus radial distance. To match the observed electron column density profile, we have fit the local electron density profile as two power laws joined at 7.8 $R_J$ (solid line). The Voyager 1 electron density profile of Bagenal, 1994 (dotted line) is shown for comparison. The local density profile integrated over the sight (dot-dash line) closely matches the observed electron column density profile.

**Figure 7**. Model derived mixing ratios as a function of radial distance. The solid lines are the UVIS results, while the dotted lines are the parameters from the Voyager-based model of Bagenal (1994). The error bars are the formal 1-σ errors obtained from the fitting algorithm.

**Figure 8.** Four selected 2-D confidence intervals of the model parameters. The contours represent the value of $\Delta\chi^2$ corresponding to the probability of finding the pair of parameters within the contour. The cross in the center of the panels represents the formal 1-$\sigma$ errors (i.e. the square root of the diagonal elements of the covariance matrix) obtained from the fitting algorithm. The formal 1-$\sigma$ bars often, though not always, underestimate the true range of possible parameter values. The upper panels show examples of two pairs of parameters that are only weakly correlated, while the bottom panels show pairs of parameters that are highly anti-correlated.

**Figure 9.** Normalized distribution functions for the Maxwellian and kappa distributions fit to the spectrum at 7.4 $R_J$. The Maxwellian distribution contains more particles than the kappa in the energy range of 5-40 eV. Consequently, model fits using a kappa distribution require a higher electron column density than those using a Maxwellian distribution.

**Figure 10.** Best-fit values of the $\kappa$ parameter versus radial distance. The solid diagonal line is the best-fit line through the values of $\kappa$. The labeled M-V 95 is the value of $\kappa$ determined from the Ulysses URAP instrument during the Io torus flythrough in 1992 [Meyer-Vernet et al. 1995]. The decrease of $\kappa$ inside of 6.5 $R_J$ may be due to line of sight projection effects.

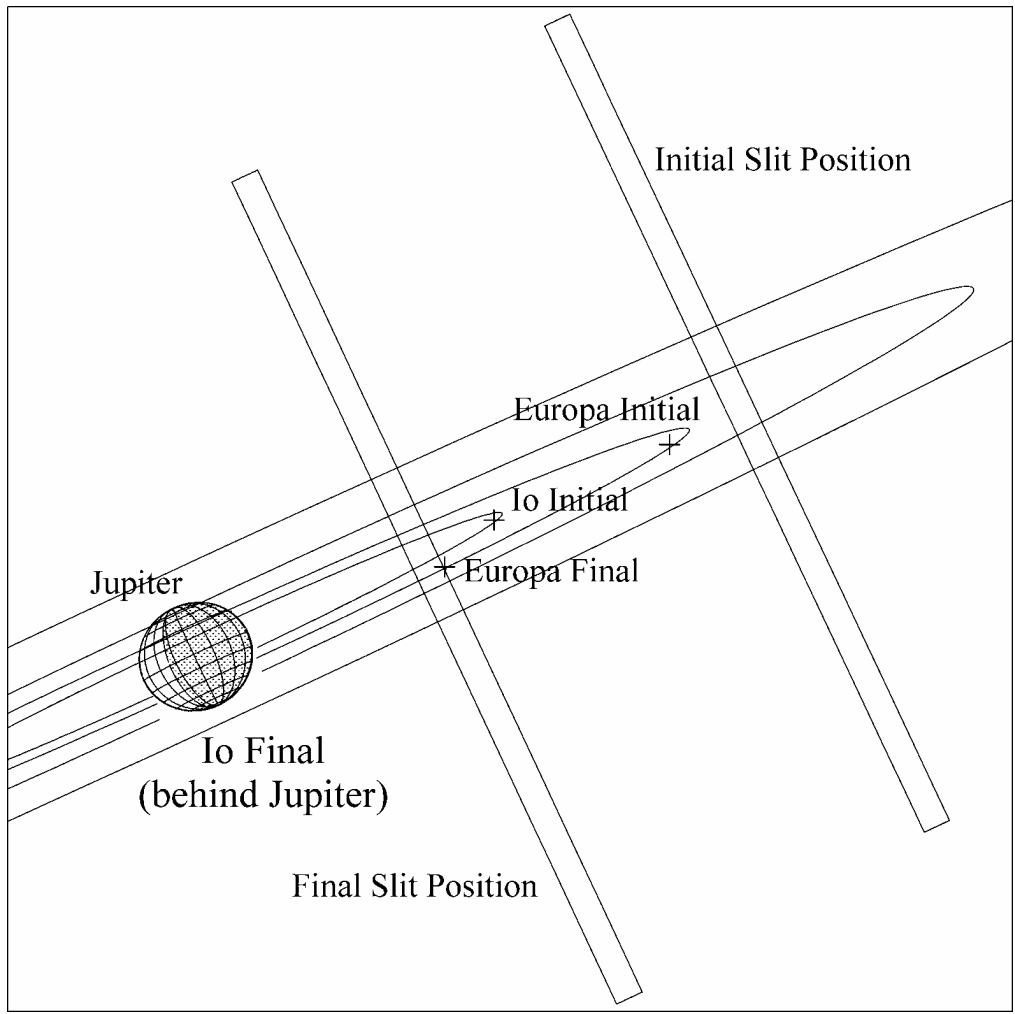

FIGURE 1. Steffl *et al.* Io Torus Radial Variations

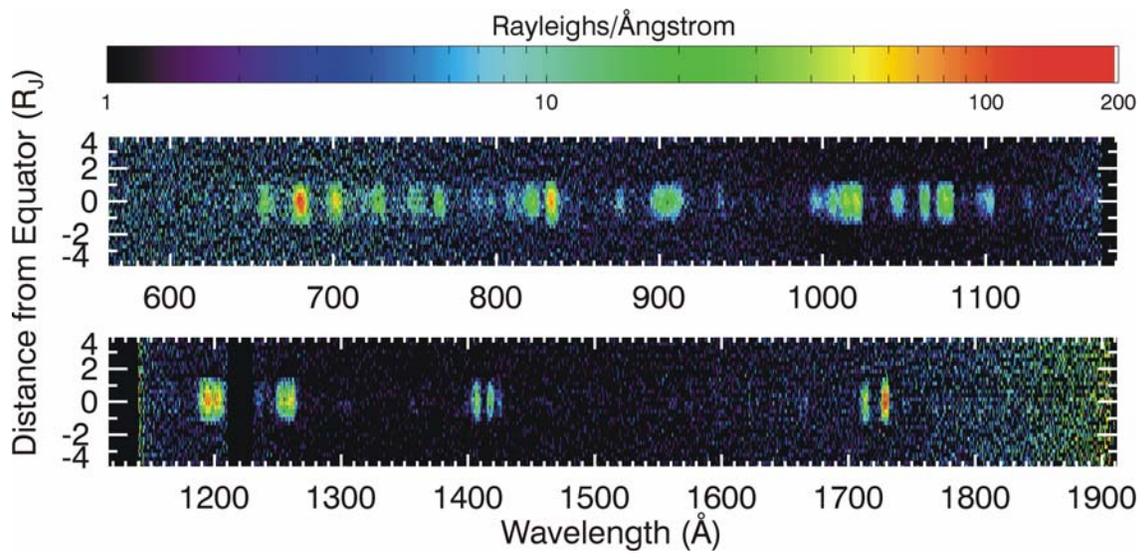

FIGURE 2. Steffl *et al.* Io Torus Radial Variations

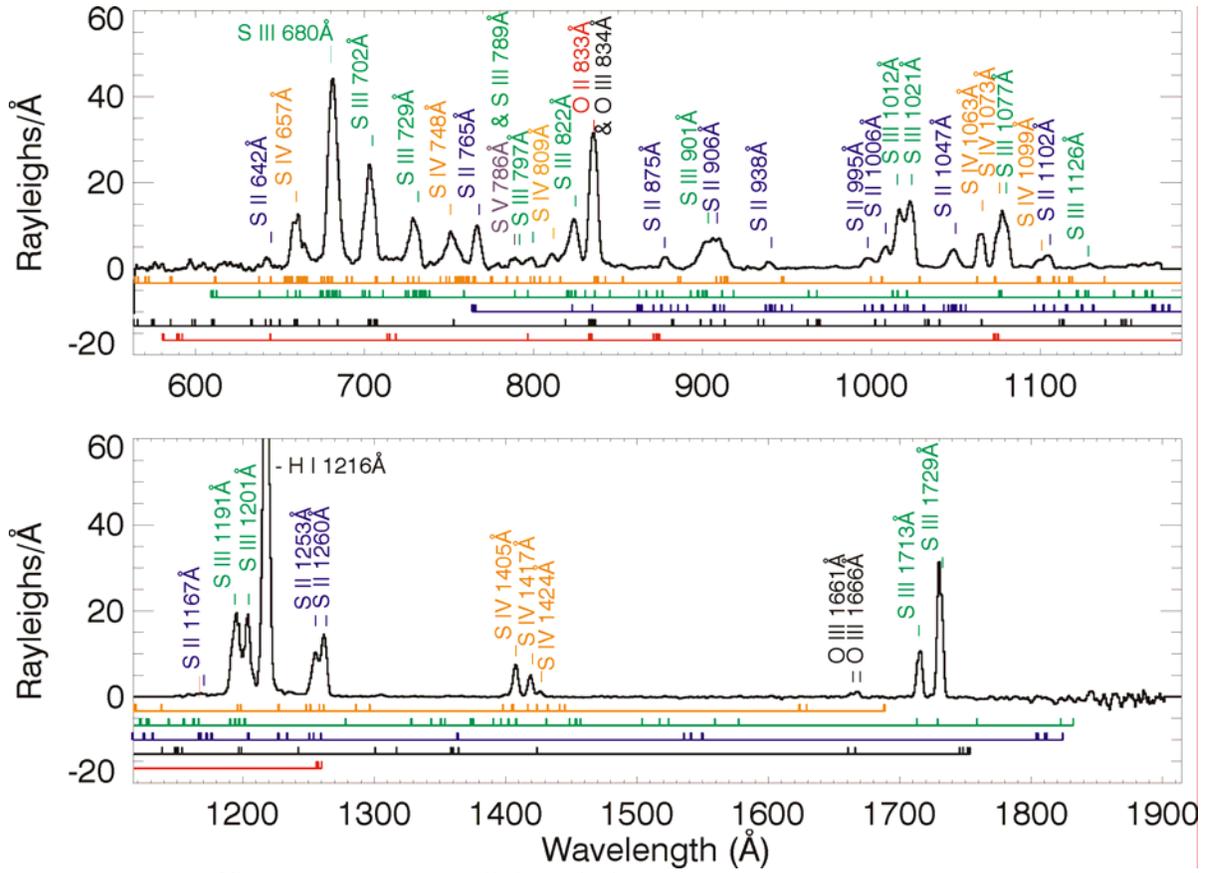

FIGURE 3. Steffl *et al.* Io Torus Radial Variations

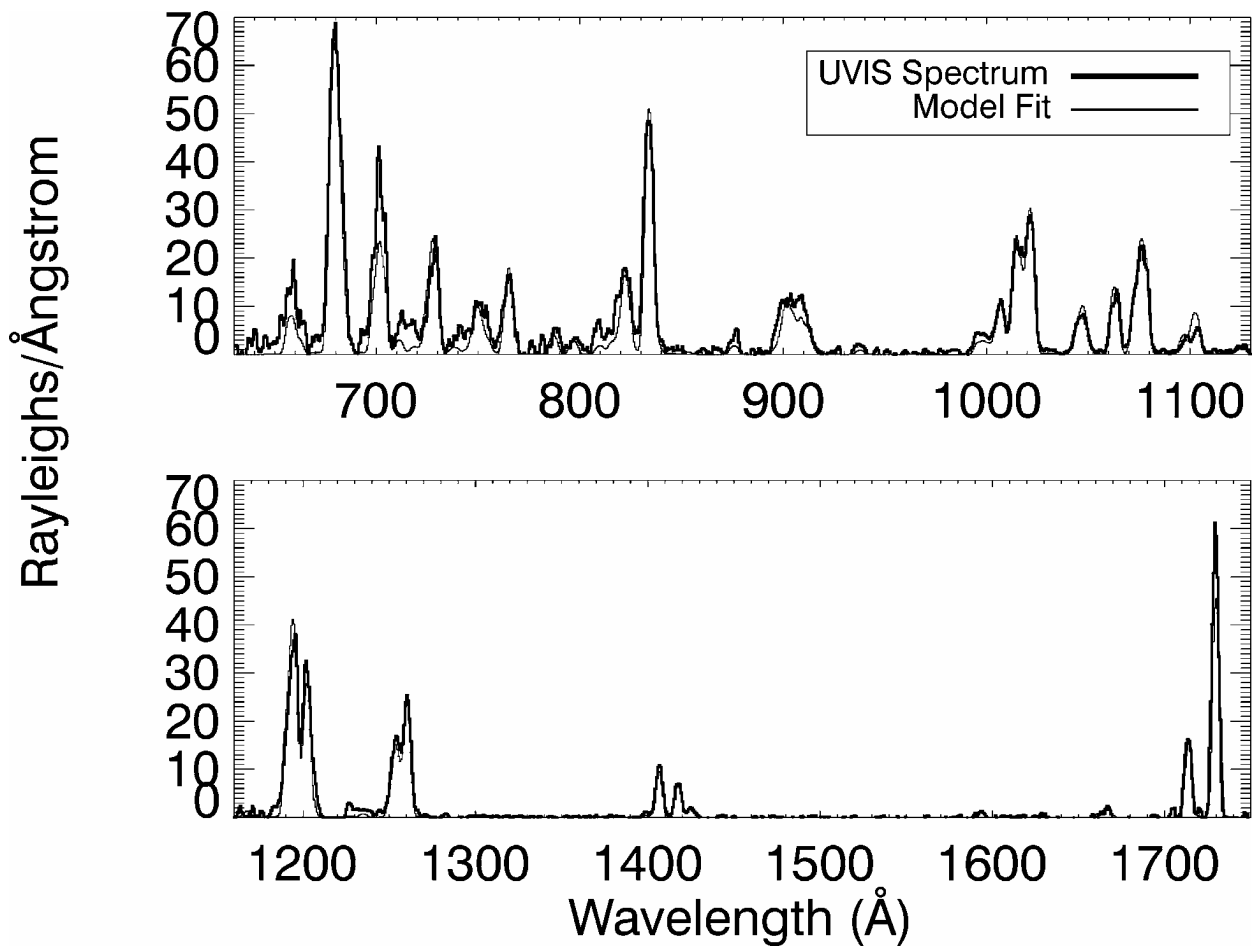

FIGURE 4. Steffl *et al.* Io Torus Radial Variations

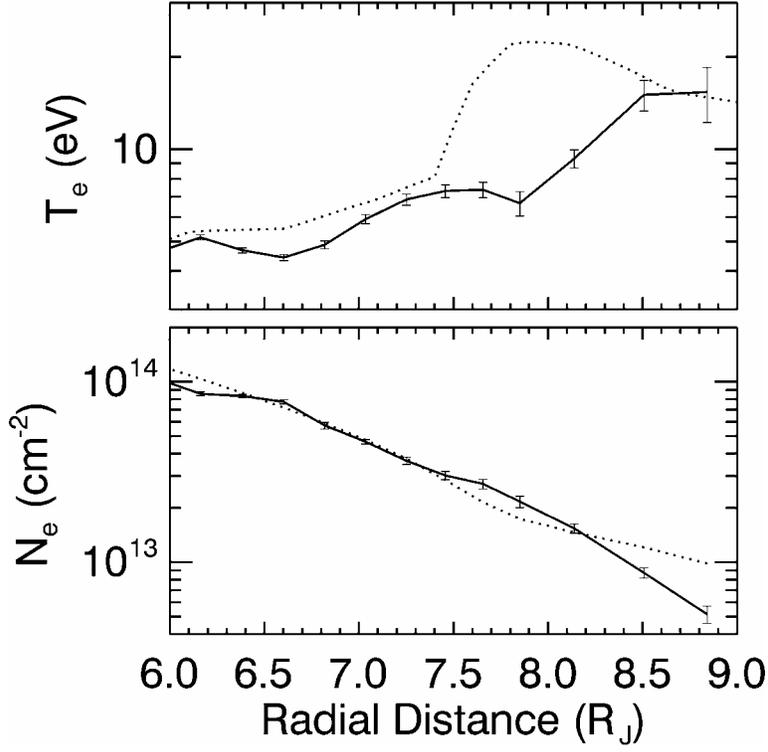

FIGURE 5. Steffl *et al.* Io Torus Radial Variations

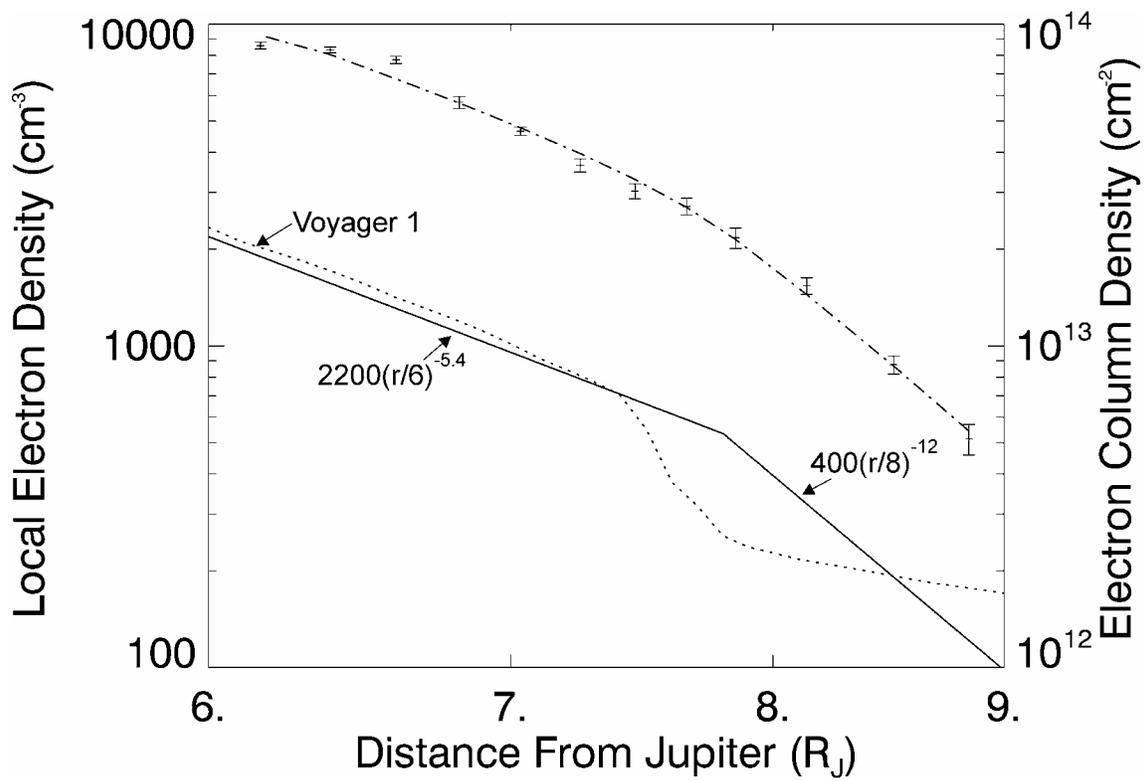

FIGURE 6. Steffl *et al.* Io Torus Radial Variations

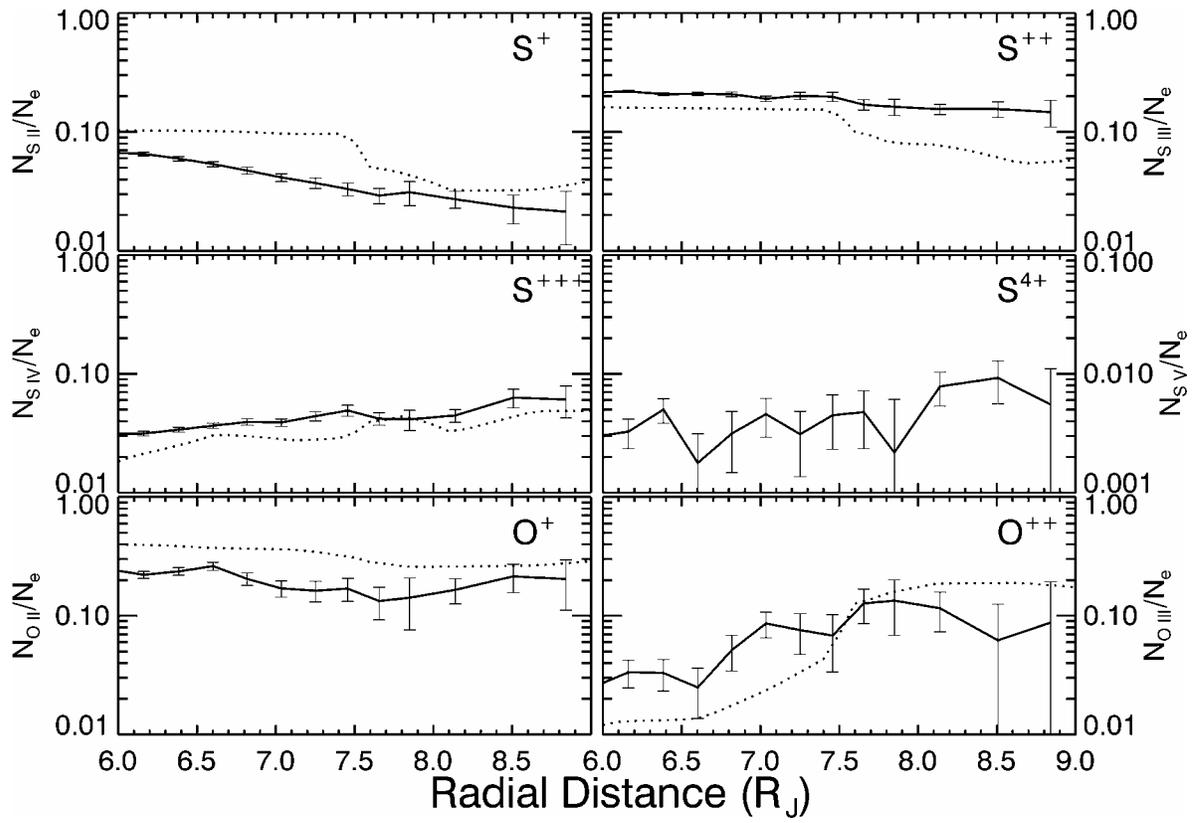

FIGURE 7. Steffl *et al.* Io Torus Radial Variations

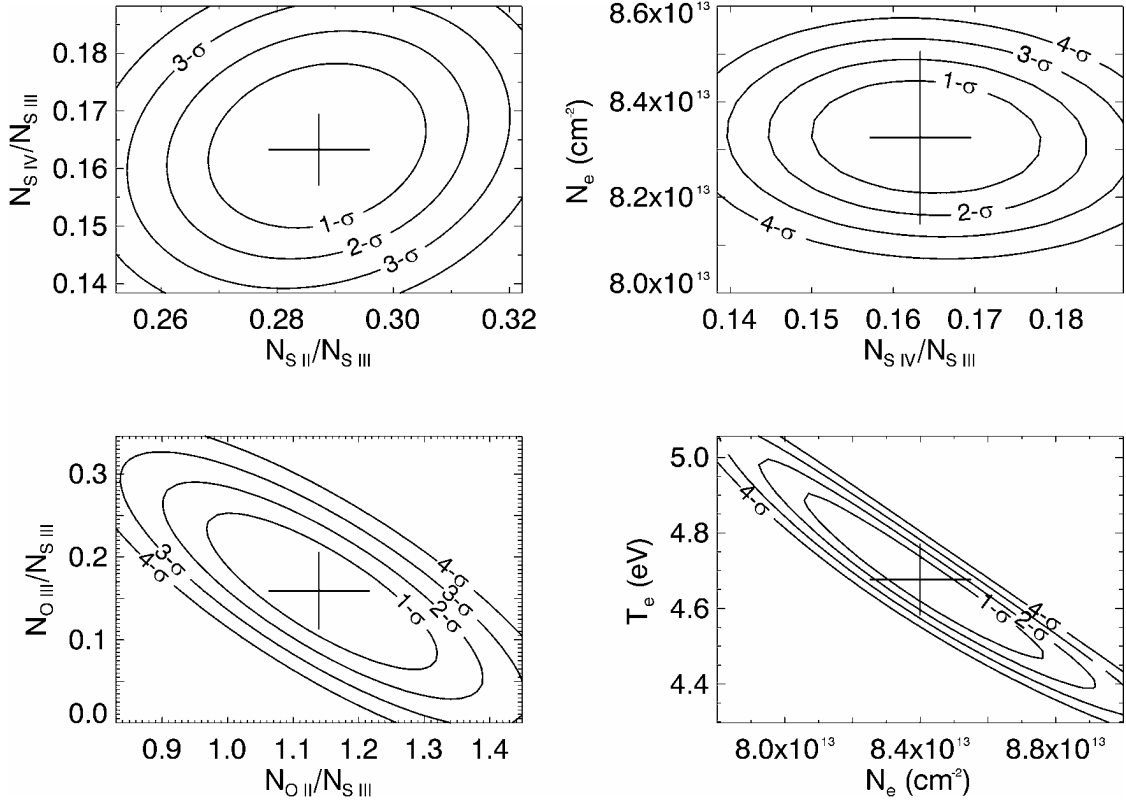

FIGURE 8. Steffl *et al.* Io Torus Radial Variations

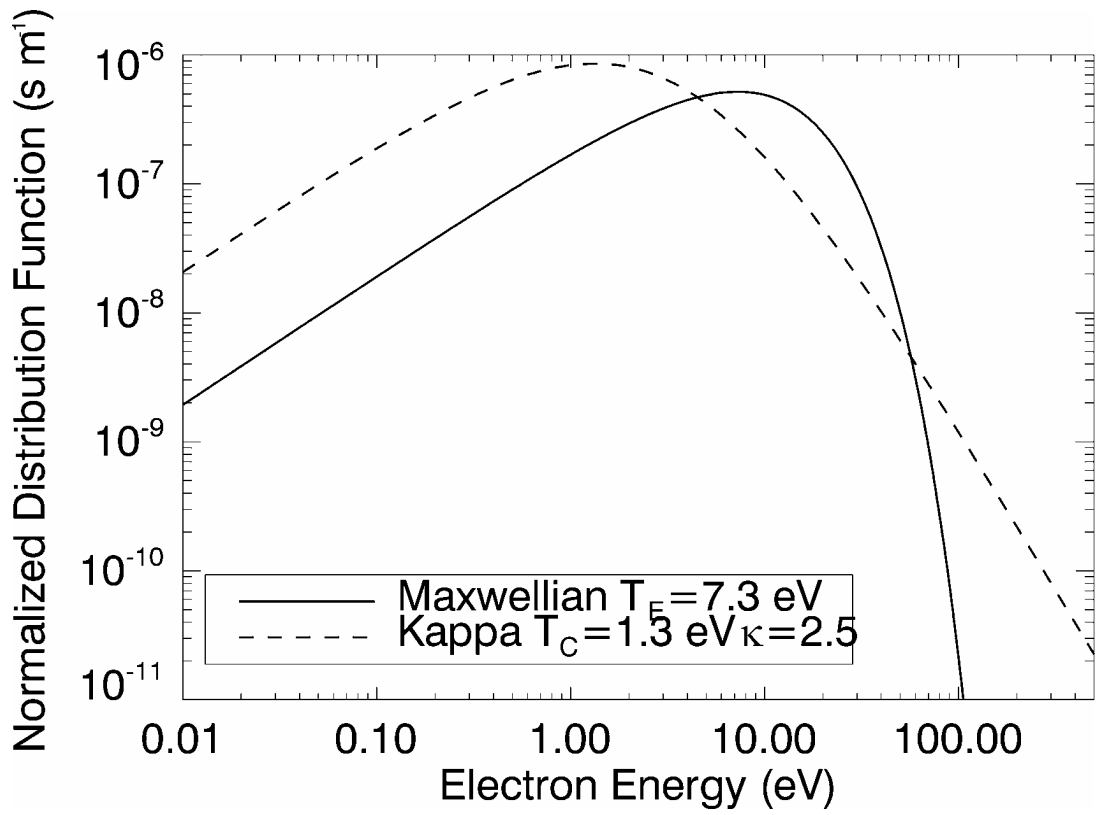

FIGURE 9. Steffl *et al.* Io Torus Radial Variations

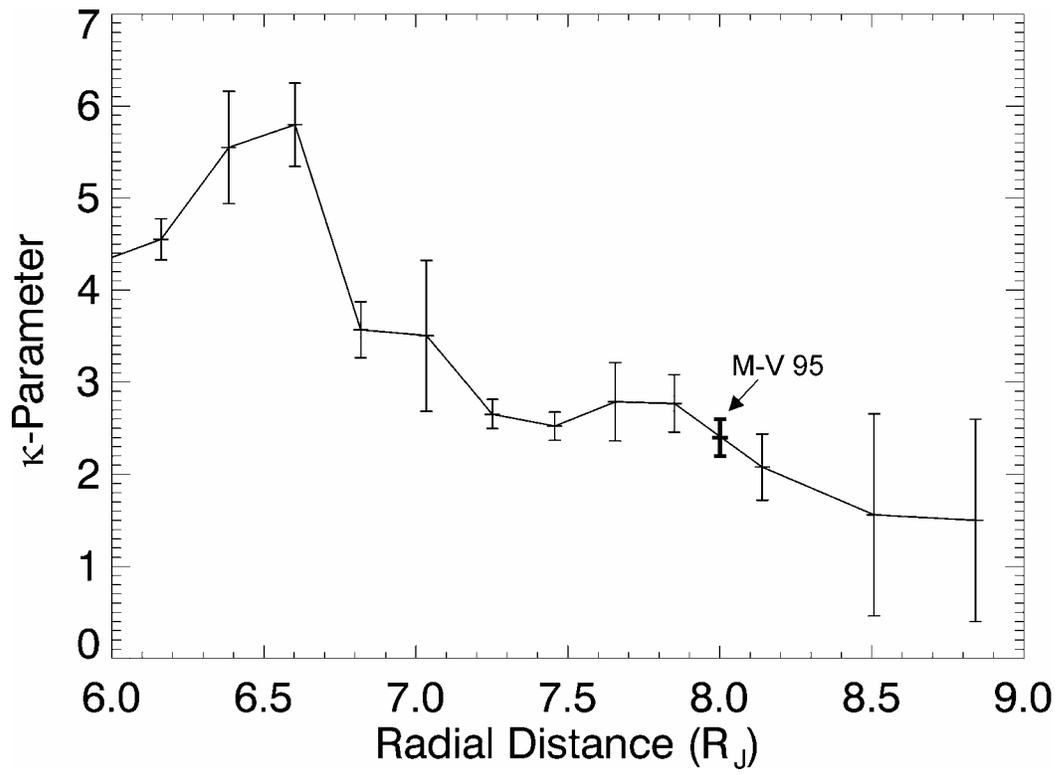

FIGURE 10. Steffl *et al.* Io Torus Radial Variations